\documentclass{article}
\usepackage{graphicx} 
\usepackage{hyperref}
\usepackage{amsmath}
\usepackage{mdsymbol}
\usepackage{mathtools}
\usepackage{dsfont}
\usepackage [english]{babel}
\usepackage [autostyle, english = american]{csquotes}
\MakeOuterQuote{"}

\title{Practical Problems of Statistical Learning}
\author{Joseph Andersen \thanks{University of Oregon - Undergraduate Student}}
\date{June 2023}

\usepackage{parskip}
\begin{document}
\maketitle

\begin{center}
\emph{This paper is submitted to the University of Oregon Department of Economics, in partial fulfillment of the requirements for departmental honors.}
\end{center}

\vspace*{\fill}
\begin{abstract}
Statistical models have seen a significant rise in popularity in recent years. Despite their undeniable success in various industry use cases such as sabermetrics, investment portfolio management, and artificial intelligence, there has been immense debate about the value of results produced by statistical methods. This paper focuses on presenting the common issues practitioners have when implementing statistical learning models, and why these issues make it difficult to interpret results produced by such methods. 
\end{abstract}
\vspace*{\fill}

\newpage

\tableofcontents


\newpage
\section{Acknowledgements}
I would like to thank Professor Edward Rubin for his helpful guidance throughout the writing and research process of this paper. I would also like to thank ChatGPT for answering many of my technical questions. 


\section{Introduction}
\subsection{The Death of Theory} 
The past couple decades have seen an unstoppable march away from principled and deductive thinking, to a data-centric approach whenever a problem presents itself. In sciences concerned with complex social phenomena, where causal relationships resemble a tangled web, an empirical approach is often the quickest way to finding answers to a problem. In 2008 journalist Chris Anderson published a highly provocative article titled \emph{The End of Theory: The Data Deluge Makes the Scientific Method Obsolete} \cite{A08}. The article argues that with enough data and compute it will be more efficient to abandon theory in favor of models based on the frequency of past events. Rather than trying to understand how a single variable affects natural phenomena, we can simply target it as an outcome variable and induce what will happen. Anderson's article ignited a wave of debate about whether the scientific method is necessary in an era of cloud compute and data abundance. On one extreme, data scientists argue that statistical methods can replace the scientific method and called for the death of theory. On the other hand, epistemologists and academics who had dedicated their lives towards understanding social phenomena, argue that statistical methods are trivial, and don't contribute to actual scientific knowledge. They proclaim that such methods can only approximate the physical laws they attempt to discover.

This paper seeks to provide a broad overview of what problems prevent statistical learning from producing scientific knowledge. In doing so, we hope to better understand the extent to which statistical learning models are valuable. Out of scope of this paper are the social and political issues of how data is collected. Although this list is expansive, our paper is limited to practical issues of implementing statistical learning. 

\subsection{Statistics}
Statistics is a field of mathematics that primarily aims to transform raw data into refined information. Statistical inference is the process that allows scientists to induce knowledge from a sample onto a population. Although used in many contexts, this paper is only focused on looking at statistical inference in the causal inference and prediction settings. 

\subsection{Causal Inference}
The first setting of statistical learning is causal inference. This discipline focuses on developing mathematical models that allow us to understand the causal relationship between an independent and dependent variable. The fundamental problem of causal inference is that for each individual $i$, we only observe outcome $Y_{i}(1)$ or $Y_{i}(0)$. As time is linear, we are not able to observe the counterfactual of an individual. Rather, we are only able to observe what would happen if an individual is treated or not treated. Due to the presence of covariates and confounders, this means a direct comparison between treated and untreated individuals is typically not equal. As a result it becomes difficult to extract the causal effect of a variable.

In order to address this problem, researchers use randomization and the law of large numbers to generate a control and treatment group. Since observations are randomized we are able to reasonably assume that biases are equally distributed between the groups, and are effectively able to create a counterfactual. Approaches to randomization generally fall into experimental or quasi-experimental methods. 

Experimental methods, such as randomized controlled trials, randomize participants into control and treatment groups. If the sample size is large enough and the randomization process is sound, researchers do not need to worry about bias and thus the potential outcome $Y(0)$ or $Y(1)$ is independent of the treatment itself. As a result the average treatment effect can be correctly obtained with the regression model (1).

\begin{center}
    \begin{equation}
        Y_{i} = \alpha + \tau D_{i} + \epsilon_{i}
    \end{equation}
\end{center}

In this model $Y_{i}$ is the outcome, $\alpha$ is the intercept, $\tau$ is the average causal effect of the treatment on the outcome, $D_{i}$ is the treatment variable, and $\epsilon_{i}$ is the error of the model. 

Pure experimental methods are often impractical due to economic issues, ethical concerns, or attrition. This is especially true for econometricians who study macroeconomic phenomena. In this situation, researchers resort to quasi-experimental methods in which they attempt to \emph{approximate} true randomization.

The first type of quasi-experimental method is \emph{selection on unobservables}. In this approach, researchers attempt to control for bias by carefully selecting what observational data they use in their model. This approach boils down to finding a data generating process in which eligibility for treatment is continuous, and the variation of the unobservable characteristics is smooth at the discontinuity where observations become eligible for treatment. The principal assumption in selection on unobservables, is that the discontinuity creates random variation in which observations are treated. The data selection process in this method is known as the \emph{identification strategy} and often requires an extremely deductive process. The validity of this assumption may be tested via a manipulation test or covariate smoothness test. This approach allows researchers to approximate the randomization of an RCT experiment, and then extract the average causal effect of a treatment by implementing research designs such as a regression discontinuity (2) or difference-in-differences approach (3). 

\begin{center}
    \begin{equation}
        Y_{i} = \alpha + \tau D_{i} + \beta_{1} (X_{i} - c) + \beta_{2} (X_{i} - c)D_{i} + \epsilon_{i}
    \end{equation}
\end{center}

\begin{center}
    \begin{equation}
        Y_{i} = \alpha \mathds{1}(post) + \gamma \mathds{1}D_{i} + \tau \mathds{1}(post) * \mathds{1}D_{i} + \epsilon_{i}
    \end{equation}
\end{center}

In a local linear regression discontinuity model (2), the $\beta_{1}$ coefficient controls for the slope estimate before the discontinuity $c$. $\beta_{2}$ controls for the slope estimate after the discontinuity in which observations become eligible for treatment. The estimate $\tau$ is the average causal effect of the treatment assuming perfect compliance. If non-compliance is suspected, researchers may implement further controls such as a fuzzy regression discontinuity design. 

In the difference-in-differences model (3), the dummy variable $post$ indicates whether the observation was recorded before or after the treatment was implemented. Thus the average causal effect $\tau$ is found by looking at the interaction between whether the observation is in the treatment group and whether the treatment has been applied yet.

Oftentimes the treatment in an experiment is endogenously determined. This happens when the treatment is correlated with the error $Cov (D_{i}, \epsilon_{i}) \neq 0$. Informally, this means the assignment of a treatment is not randomly determined (exogenous) and is confounded by factors that influence both the treatment and outcome. To recover the causal effect a researcher must use an Instrumental Variable (IV) to isolate exogenous variation of the treatment variable. IV work by finding a valid instrument $Z$ that is correlated with the treatment, yet uncorrelated with the error term $Cov (D_{i}, Z_{i}) \neq 0$. Then researchers determine whether the instrument meets the exclusion restriction $Cov(Z_{i}, \epsilon_{i}) \neq 0$. Informally, the exclusion restriction requires $Z$ only affects $Y$ through $D$. This condition is not testable. 

The IV estimate is then found by a two stage process. The first step is known as the \emph{first stage regression} where $D \sim Z + X$. In the second step, researchers solve the \emph{reduced form regression} by $Y \sim Z + X$. The IV estimate of the causal effect of $D$ on $Y$ is then found by dividing the reduced form by the first stage (4). This method is equivalent to the Complier Average Causal Effect (CACE) which is found by dividing the intent-to-treat by the compliance rate. However, the power of Instrumental Variables comes from the fact that knowledge of the compliance rate is not needed. This allows researchers to extract causal effects in quasi-experimental and uncontrolled scenarios. 

\begin{center}
    \begin{equation}
        \frac{Y \sim Z + X}{D \sim Z + X}
    \end{equation}
\end{center}

An example of how instrumental variables in combination with a quasi-experiment research design allows researchers to extract an average causal effect is in \cite{E+17}. Ebenstein et al. used China's Huai River Policy as part of their identification strategy to determine the impact of air pollution on life expectancy. This policy heavily incentivized the use of coal for indoor heating on the Northern side of the river, while discouraging it on the Southern side. Central to identifying the causal effect, was the issue that life expectancy is endogenous to air pollution. Wealthier families that can afford better health care may have self selected into the Southern side of the river. Because of the assumption that living North of the river is correlated with pollution caused by coal heating, the authors were able to use the geographical location of observations as an instrument. This allowed the authors to create a regression discontinuity design to understand the causal effect of airborne particulate matter on life expectancy. 

As a last resort when a researcher is unable to obtain a valid identification assumption, a \emph{selection on observables} quasi-experimental approach may be used. In this situation the outcome of the treatment is not suspected to be independent of the treatment itself $Y_{i}(1), Y_{i}(0) \upvDash D_{i}$. Thus, finding the causal effect requires the the addition of controls. This is known as the Conditional Independence Assumption $Y_{i}(1), Y_{i}(0) \upvDash D_{i} \mid X_{i}$. Controls may be implemented either by matching methods, or if the researcher believes they know the functional form of the synthetic control, they may implement a control $X_{i}$ directly into the regression (5). 

\begin{center}
    \begin{equation}
        Y_{i} = \alpha + \tau D_{i} + \gamma X_{i} + \epsilon_{i}
    \end{equation}
\end{center}

This brief overview demonstrates how experimental and quasi-experimental research designs, in combination with instrumental variables, allow researchers to isolate the average causal effect $\tau$ of a treatment variable $D_{i}$ on an outcome $Y_{i}$. In the next section we will provide a brief overview of how prediction modeling works, before discussing the practical problems of implementing both types of these models. 

\subsection{Prediction}
In contrast to causal inference the prediction setting is solely concerned with the model's ability to estimate the outcome $Y_{i}$. Because of the disregard for causality, researchers are willing to accept the use of correlative variables if they have predictive power. 

The methodology of a statistical engineer was almost indistinguishable between prediction and causal inference tasks up until the early 2000s. In \cite{B01}, Breiman details how the approach to prediction was traditionally through \emph{data modeling}. Through observations, engineers deduced the functional form of the relationship they were approximating, selected a parametric model with the correct form, and trained the model until they calculated parameters that reliably estimated the outcome variable. The proliferation of ridge regression exemplifies the data modeling approach. In order to prevent overfitting, a central problem in prediction, researchers simply added a penalty term to the linear regression model.

\begin{center}
    \begin{equation}
        \min_{\beta^{R}} \sum_{i = 1}^{n}(Y_{i} - \hat{Y_{i}})^2 + \lambda \sum_{j = 1}^{p} \beta^2_{j}
    \end{equation}
\end{center}

Although functionally different from (1), the concepts used in ridge regression prediction (6) are extremely similar to those in causal inference. Training this model returns parameters that minimize the squared error of a multiple linear regression model $\hat{Y_{i}}$, whose functional form was decided \emph{a priori} to training. Parametric models such as lasso regression were ubiquitous throughout the data modeling era of prediction. By the early 2000s however, the prediction community began its shift to \emph{algorithmic models}. In \cite{B01}, Breiman writes how algorithmic modeling did not come from the statistics community but rather from a new discipline called machine learning. 

In contrast to data modeling, the learners used in algorithmic modeling do not have strict parameterization. Algorithmic models were designed to allow the learner to find the functional form of the relationship it is approximating. The use of highly flexible algorithms that can determine their own functional form, in combination with a willingness to use correlative instead of causal data, resulted in the deductive process of causal inference and data modeling being stripped out in prediction. Many of the state of the art prediction models such as OpenAI's GPT-4 have hundreds of billions of parameters. This is in stark contrast to the causal inference models used by econometricians, which rarely have more than a couple thousand parameters in the most extreme cases of panel data and fixed effects models. 

Algorithmic models include neural networks, k-nearest neighbors, decisions trees, and a subset of support vector machines. Although these models are more flexible and can approximate more complicated relationships, they are not without downsides. The training process for algorithmic models is extremely compute heavy and inefficient. This is because the complexity of the models makes them non-differentiable. As a result, models must estimate parameters using optimization algorithms such as gradient descent. These algorithms solve for partial derivatives and use the resulting vector to step the parameters towards a local minima. This process is not only time and resource intensive, but can cause the model to become stuck in sub-optimal minima. Algorithmic models also require much more training data as their flexibility makes them prone to over-fitting. However, these problems are typically insignificant relative to algorithmic models performance over data models. 

This overview demonstrates that current machine learning algorithms allow researchers to build highly predictive models as the models learn the functional form of the relationship themselves. In the next section, we discuss the practical problems of implementing statistical learning in both the causal inference and prediction setting. 

\section{Universal Problems of Statistical Learning}
\subsection{Internal vs. External Validity}
Statistical inference is considered to have \emph{internal validity} if the results are valid for the sample population. In the causal inference setting, this would mean that the coefficient $\hat{\beta_{x}}$ is equivalent to the average causal effect of $\hat{x}$ on $\hat{y}$ in the sample. In the prediction setting, internal validity would mean that when the model is given the observed variables for $y_i$, its parameters return an estimate $\hat{y_i}$ that is close to the actual outcome $y_i$.

On the other hand, statistical inference is only considered to have \emph{external validity} if the model can generalize to a broader population. In causal inference, this means that the treatment effect of $x$ on $y$ is applicable to populations and settings outside of the sample. The reason that external validity is extremely difficult to obtain in causal inference, is because of how a different setting introduces new sets of parameters and factors. Furthermore, causal inference typically focuses on cardinal rather than ordinal effects. This means that for a model to have external validity, the underlying distribution of the new population must be equivalent to the sample population it was modeled upon. For a prediction model to have external validity, it must be able to predict the outcome $y_{i}$ given a vector of variables it was not trained on. 

The reason external validity is so hard to achieve in causal inference, is because participants are often not representative of the broader population. Even in experimental designs such as randomized controlled trial, there are often many problems of finding suitable samples. In \cite{J+09} the authors surveyed a number of RCT primary care experiments, and found that participants in clinical studies were often not representative in actual primary care settings. This is problematic because clinicians must have confidence that a drug will have the same effect on a patient, as it did on the participants in clinical trials. If there is significant heterogeneity between a patient versus those participating in clinical trials, it is likely that the results will be different. 

The problem of external validity in the causal inference setting is only further exacerbated by an increasing use of quasi-experimental methods. Recall the selection on unobservables and selection on observables experiments discussed in section 2.3. These quasi-experimental designs rely on natural data generating processes as opposed to true randomization. As a result, the covariates are specific to the geographical area that was studied, and often do not translate to broader populations. When assumptions in these experiments are further relaxed, such as non-compliance, external validity is further broken down by methods that attempt to address such issues. 

This problem has caused the value of empirical economic experiments to be hotly debated. The notion of a local average treatment effect (LATE) was introduced to highlight how effects in these studies are often "local" to the subsample studied and can rarely generalize to a broader population. In \cite{I10} Guido W. Imbens provides an overview of how many notable theoretical economists struggle to place value on such estimates. Imbens writes that depending on the setting and subsample studied, there is significant heterogeneity in the estimands of empirical research. As a result a single estimate is unlikely to be useful for informing policy. Instead, researchers can obtain informative estimates by combining several studies based on multiple populations and settings. Although researchers are often unable to find the true population mean, the collection of multiple local average treatment effects allows for a conglomerate of experiments with individually weak external validity, to provide value to policy makers and researchers. 

External validity in the prediction setting is very similar and yet very different to causal inference. Like causal inference, prediction faces the problem of gathering a training set that has the same distribution as the broader population. If a learner is trained on data that is unrepresentative it will perform poorly. However, this is not the only problem of prediction. Researchers must also be aware of underfitting and overfitting. 

Machine learning algorithms are highly flexible and can fit non-parametric relationships in high dimensional spaces. Without restraining this flexibility, the algorithm will simply learn to interpolate every observation it is trained on. This phenomenon is known as overfitting, and is problematic because it results in the model learning the statistical noise of the relationship it is approximating, rather than the underlying relationship itself. When a model overfits it performs extremely well in-sample, but extremely poorly out-of-sample. In contrast when the model's flexibility becomes too restricted, the model underfits the relationship. For example consider a parametric model, such as a linear regression, attempting to approximate a non-parametric relationship. 

To guard against the problem of overfitting and underfitting, machine learning algorithms rely on two approaches. The first is regularization which introduces a cost on the flexibility of a model. This can be an explicit cost such as lasso and ridge regularization. On the other hand, regularization can happen via randomization such as in drop out or feature subsampling. To achieve the correct amount of regularization, researchers will partition their data set so that they may \emph{cross validate}. What this means is that they train and adjust the amount of regularization on one portion of the data, and then test it out-of-sample on the remaining data.

These examples highlight the problem of external validity in causal inference and prediction. Although there are methods to address generalization in both settings, it is important to not place too much value on any single model, even if it has high internal validity. By creating models on different sub populations, and observing the heterogeneity in results, we are better able to approximate the population mean. 

\subsection{The Importance of Context}
Statistical learning methods also face severe interpretability problems if the correct preprocessing and deduction processes are not undertaken. Although the exact effects of these problems differs in magnitude between causal inference and prediction, the problem can be boiled down to the simple fact that without context and interpretation, data is meaningless. 

A commonly used example to explain the importance of context in statistical analysis, is a car driving down a road with varying slope. Consider a driver who is going down this hill and wants to maintain a constant speed. If the driver is skilled, pressing the brake will have no correlation with the speed despite the fact that it has a causal effect on the constant speed. 

An observer with no knowledge of the causal relationships between these variables may conclude that there is no relationship between the brake and the speed. In causal inference this is known as endogeneity, where the treatment effect of the break is correlated with the error term. If the observer attempted to regress the speed on the break pressure and hill height, they would encounter multicollinearity and receive biased estimates. This example demonstrates how understanding a systems mechanics is vital to properly interpreting causal inference estimates.

Context is equally important in prediction. The example of the car and the hill demonstrates that a lack of correlation does not mean a lack of causation. As machine learning algorithms are trained on correlations, it is important to note that a learner with poor predictive accuracy does not mean that causal relationships are not present. Alternatively, just because a learner has high predictive accuracy does not mean that the correlation is not spurious. 

Regardless of whether one is in the causal inference or prediction setting, the use of statistical learning without context is perilous. Simply throwing data at an algorithm will not make up for an improper understanding of how a system works. 

\subsection{Curse of Dimensionality}
All statistical learners face the curse of dimensionality. This problem encapsulates the challenges that arise from dealing with high-dimensional data. In causal inference, the curse of dimensionality applies when the model has many terms (also known as dimensions). Consider the following models:

\begin{center}
    \begin{equation}
        Y_{i} \sim \alpha + \tau D_{i} + \epsilon_{i}
    \end{equation}
\end{center}

\begin{center}
    \begin{equation}
        Y_{i} \sim \alpha + \tau D_{i} + \beta_{1} X_{1} + \beta_{2} X_{2} + ... + \beta_{100} X_{100}+ \epsilon_{i}
    \end{equation}
\end{center}

The parameters of an ordinary least squares regression are only differentiable when the model has fewer terms than data observations. Model (6) only requires three observations to find an estimate for $\tau$ while model (7) requires one hundred three observations. However, this is not the only problem. Increased complexity of the model causes data to become sparse which may cause biased estimates. This is especially prevalent when the model must control for observable characteristics.

Matching involves pairing treatment observations to control observations based on whether they share the same covariates. If the treatment and control observations share the exact same observable characteristics, we can assume $Y_{i}(1), Y_{i}(0) \upvDash D_{i} \mid X_{i}$. However as more and more dimensions are added, it becomes increasingly unlikely exact matches will be found. When this happens researchers must result to non-exact matches and use approaches such as k-nearest neighbors to find approximate matches. Alternatively, controls may be implemented using propensity score matching. In this method observations are matched based on whether they have the same conditional probability of receiving treatment. Although this approach allows researchers to reduce the amount of dimensions of their model, it introduces bias, can skew estimates, and introduces additional causality assumptions. Implementing the controls directly into the regression such as in (4) does not solve this problem either as it adds even more terms to the model. 

In prediction, the curse of dimensionality presents a different set of problems. Recall the problem of finding a local minima discussed in section 2.4. Adding dimensions to your model makes computing the partial derivatives more and more compute heavy. Furthermore, adding more dimensions to your model, while retaining the same amount of observations, makes it more likely that your model will over fit. Dimensionality reduction techniques such as principal component analysis (PCA) and uniform manifold approximation and projection (UMAP) were invented to overcome these problems, however introduce a host of their own problems. 

In summary, the curse of dimensionality is a universal problem of statistical learning. Understanding how potential solutions to the curse of dimensionality biases results is paramount to building useful models. 

\section{Problems Unique to Causal Inference}
\subsection{Small Problems, Small Significance}
Causal inference heavily constrains economists to the type of problems that may be studied. Economists and other social scientists are concerned with understanding social phenomena. These problems are often enormous in scope, and make it difficult to run experiments on. For example, what is the effect of an authoritarian versus democratic government on the GDP of a country? 

The scope of these questions have been a large reason that most economic work has historically been concerned with structural models rather than empirical ones. In \cite{I10} while advocating for the continued use of econometric methods, Imbens presents the arguments raised in \cite{D09} and \cite{H+09}. Deaton, Heckman, and Urzura state that economic research has become excessively experimental. They advance their argument by presenting the difficulty of interpreting local average treatment effects, but also state that empirical methods handcuff economists from answering interesting economic questions. In \cite{H+09} the authors write, "Proponents of [instrumental variable approaches] are less ambitious in the range of questions they seek to answer. The method often gains precision by asking narrower questions. The problem that plagues the IV approach is that the questions it answers are usually defined as probability limits of estimators and not by well formulated economic problems". 

These necessary conditions restrict researchers using statistical learning methods from looking at broader questions. Distilling a question of interest into an independent and dependent variable, and then isolating exogenous variation of a treatment variable greatly constrains the set of all available problems. As a result, empirical economic work has been forced to focus on partial equilibriums that struggle to find a place in broader economic frameworks.

\subsection{Hypothesis Testing}
Causal inference has additionally faced scrutiny because of its abuse of hypothesis testing. When determining the validity of an estimated parameter, researchers calculate P-values. These allow researchers to obtain a statistical measure of the likelihood of an observed outcome, assuming the null hypothesis is false.

In \cite{G+13} Gelman and Loken give an overview of the statistical crisis in science. In the pursuit of publishing research, academics have come up with ingenious ways of producing statistical significance even when it is not present. This practice has given rise to the colloquialisms "P-hacking" and "hypothesis fishing". In \cite{B+16} the authors examined top economic journals and found an unusual distribution of statistical significant results. This is attributed to researchers who artificially inflate the statistical significance of their research. This has not only raised many questions about the credibility of the researchers involved, but also of empirical economics as a whole.

The ability to examine data in many different ways makes it possible to achieve statistical significance from pure noise. Whether it be by excluding observations or selecting certain interactions in the regression model, a statistically significant hypothesis can be fit to the data instead of the other way around. This problem is especially prevalent in research that relies on instrumental variables. As the exclusion restriction assumption is not testable, researchers may try a variety of instrumental variables until they find one that produces a statistically significant estimate. 

In response to this, many have advocated for researchers to make it easier for others to reproduce their work. By providing data sets as well as code for their methods, scientists hope to solve the statistical crisis. Alternatively, others have proposed to completely eliminate the Frequentist mindset of statistical significance. They call for researchers to report the posterior probabilities of obtaining a parameter, which may be calculated via Bayesian inference. This is in stark contrast to statistical significance which reports the probability of the data given a parameter.

\section{Problems Unique to Prediction}
\subsection{Correlation is Not Causation}
Prediction models gain their power by utilizing variables that have strong correlative power. This is problematic because such variables may lack any explainable causal relationship with the outcome variable. This approach allows researchers to build extremely powerful models, but also makes it difficult to interpret the results they produce. This is because exploiting correlative relationships to build highly predictive models is sound only as long as the distribution of data does not shift. Simply making a prediction does not affect a system or the distribution of data that the prediction was pulled from. However the purpose of making predictions is to make better decisions which themselves may affect the system. Thus, making a decision based on a prediction can be circular in some settings. It is often difficult to discern whether a decision will have a causal effect on the the prediction it is predicated upon. Although this problem is not applicable to all situations, it demonstrates the issue of relying on associative rather than causal relationships. 

The reliance on correlative relationships has also caused many to question the scientific value of prediction models. This question has especially been of interest to the Linguistics and Natural Language Processing communities. In a New York Times Op-Ed titled \emph{The False Promise of ChatGPT}, Noam Chomsky writes "[ChatGPT is] a lumbering statistical engine for pattern matching, gorging on hundreds of terabytes of data and extrapolating the most likely conversational response or most probable answer to a scientific question" \cite{C+23}. Chomsky's point is that large language models such as ChatGPT are able to generate realistic-sounding language but have no understanding of what it is producing. This sentiment has been echoed by other linguists such as Emily M. Bender, who coined the term `stochastic parrot' in reference to LLMs. 

On the other hand, computer scientists such as Peter Norvig point to the overwhelmingly dominant success of statistical models in industry use cases. In \emph{On Chomsky and the Two Cultures of Statistical Learning} Norvig writes, "the intellectual offspring of [Claude] Shannon's theory create several trillion dollars of revenue each year, while the offspring of Chomsky's theories generate well under a billion". Norvig is referencing Claude Shannon's paper \emph{A Mathematical Theory of Communication} which introduced the idea of modeling language using Markov (Markkoff) chains, a type of statistical model. Shannon's work laid the foundation for the statistical language models widely used in search engines, speech recognition, machine translation, and other applications. On the other hand, the language models developed by computational linguistics have seen very limited adoption. Rather than using probability, these models are based on logical rules and semantic parsing. In contrast to `stochastic parrots' these models are intended to have an understanding of what it is producing. Norvig's point illustrates the harsh truth. Although interpreting results generated from correlative variables may be difficult, probability-based models clearly outperform all other options. Despite the fact that logic-based models are intended to have a causal understanding of semantics, they simply can not compete with statistical methods. 

\subsection{Deep Learning}
Deep learning, a subset of machine learning networks, compounds the difficulty of interpreting correlative variables by transforming and interacting them in an extremely complicated manner. Deep learning utilizes neural networks with many hidden layers. Almost every notable artificial intelligence application including ChatGPT, AlphaGO, and Tesla's autopilot, are powered by these networks. Their power comes from the fact that each layer can capture and represent data in a different way. This lets the model build a hierarchy of transformations on the inputs, and capture even the most complex patterns.

Although highly effective, this hierarchy quickly becomes inaccessible to the human observer simply due to its sheer complexity. Attempting to understand how the model builds the hierarchy during the training process is equally frivolous. Neural networks don't work properly unless they are initialized with random weights and balances. This means that understanding how the inputs are represented and transformed from layer to layer, is typically impossible even from the start of the training process.

Researchers have made many attempts at advancing techniques that let us understand how each layer of a neural networks represents and transform inputs. Seminal work by Neel Nanda et al. \cite{N+22} trained a single-layer transformer neural network to perform the math task $(a + b) \% p$. Nanda et al. found that at first the model simply memorized all of the training examples it was given. This is not uncommon, and is an example of overfitting. However as training continued, the transformer stopped memorizing the examples and instead started to solve the task by using a complicated algorithm. Rather than learning the addition and modulo operators, the transformer built an algorithm based on trigonometric identities and discrete Fourier transforms. The development of this algorithm may be partly due to the fact that the hyperbolic functions used in the activation of each neuron are similar to waves. The results of this paper demonstrate how neural networks are prone to learning how to solve tasks in methods that are completely unintuitive to humans. Attempting to understand how a single layer transformer learned took the authors many weeks. As most neural networks used in production have orders of magnitude more layers, it becomes clear that understanding exactly how neural networks are representing data is most likely impossible. 

For most tasks, it can be argued that understanding how a neural network works is not important. However for high-risk applications, a lack of understanding can be disastrous. Consider the neural networks used in high-risk computer vision tasks such as Tesla's autopilot. Understanding how the model recognizes a person could lead to the difference in life or death. Even worse is if machine learning is used in tasks for things such as identifying criminals. A paper titled \emph{Automated Inference on Criminality Using Face Images} built a model that predicted, with an accuracy of 89.5\%, whether facial features could determine whether someone was a criminal \cite{W+17}. Although the positive and negative instances used to train the model looked identical to the human observer, the learner picked up biases in both groups. In one group, most of the samples wore white collars while the other group did not. This may have reflected the light onto the samples faces, which while not observable to the naked eye, was picked up by the learner. Critics of the paper also argued that the criminal photos may have also contained micro-expressions due to the sample's recent incarceration. 

Computer scientists have long been aware of the expression `garbage in, garbage out'. However, the difficulty in understanding how deep learning networks represent data makes it much more difficult to determine whether training data really does contain biases. Without an understanding of how the data is being represented, racist, sexist, or other biases can make their way into production systems and have extreme consequences. Without a clear understanding of how data is represented in neural networks, it becomes difficult to understand the robustness of the model, and to what degree we can trust the predictions they produce. 

\newpage
\section{Conclusion}
Statistical learning has surged in importance over the past few years and become one of our most pivotal tools in our decision-making processes. Policymakers, pharmaceutical scientists, and advertisers have progressively embraced the utilization of causal inference to make informed decisions. Prediction continues to play a more and more important role in our lives as well. Deep learning has made immense breakthroughs in predicting protein structures, self-driving cars, and natural language processing. Other machine learning methods such as decision trees, have helped develop fraud detection systems and credit risk systems. Statistical learning's growing importance in our lives shows no signs of slowing.

Chris Anderson's \emph{The Death of Theory} envisions a future where we have no need for the scientific method. While the expanding influence of statistical learning may make it seem like this future is not far away, it is important to realize that these methods are not a silver bullet. Problems such as external validity, interpretation of independent variables, and the curse of dimensionality exist for both causal inference and prediction models. Additionally, both settings have an additional set of unique problems. It is crucial to acknowledge how these issues may prevent correct implementation of these methods, and how they may effect our interpretation of their results. Recognizing this allows us to create better yet still imperfect models. Additionally, staying cognizant of these issues allows us to make better value judgements of the results produced by statistical models, and use these models to their fullest extent.  
\newpage


\end{document}